\documentclass[a4paper,11pt]{article}
\pdfoutput=1 
 \usepackage{jcappub}
\usepackage{amsmath,amssymb}
\usepackage{bm}
\usepackage{xcolor}
\usepackage{graphicx}
\usepackage{enumitem}
\usepackage{geometry}
\usepackage{xspace}
\usepackage{pdflscape}
\usepackage{placeins}
\usepackage{epstopdf}
\usepackage{comment}
\usepackage{braket}
\usepackage{subfig}

\makeatletter
\gdef\@fpheader{}
\g@addto@macro\bfseries{\boldmath}
\makeatother

\let\oldsqrt\sqrt
\def\sqrt{\mathpalette\DHLhksqrt}
\def\DHLhksqrt#1#2{%
\setbox0=\hbox{$#1\oldsqrt{#2\,}$}\dimen0=\ht0
\advance\dimen0-0.2\ht0
\setbox2=\hbox{\vrule height\ht0 depth -\dimen0}%
{\box0\lower0.4pt\box2}}

\newcommand{\sss}[1]{{\scriptscriptstyle{#1}}}
\newcommand{\boldmathsymbol}[1]{{\ensuremath{\boldsymbol{#1}}}}

\newcommand{\uPl}{\mathrm{Pl}}

\newcommand{\usssPl}{\sss{\uPl}}

\newcommand{\Mp}{M_\usssPl}

\newcommand{\beq}{\begin{equation}}
\newcommand{\eeq}{\end{equation}}
\newcommand{\bea}{\begin{equation}\begin{aligned}}
\newcommand{\eea}{\end{aligned}\end{equation}}

\newlength{\wsingfig}
\newlength{\wdblefig}
\newlength{\wquadfig}
\newlength{\wtriplefig}
\newcommand{\Eq}[1]{Eq.~(\ref{#1})}

\newcommand{\Fig}[1]{Fig.~{\ref{#1}}}

\newcommand{\Sec}[1]{Sec.~\ref{#1}}

\title{Primordial black holes in loop quantum cosmology: The effect on the threshold}

\author[a]{Theodoros Papanikolaou}
 
\affiliation[a]{National Observatory of Athens, Lofos Nymfon, 11852 Athens, 
Greece}

\emailAdd{papaniko@noa.gr}

\abstract{ 
Primordial black holes form in the early Universe and constitute one of the most viable candidates for dark matter. The study of their formation process requires the determination of a critical energy density perturbation threshold $\delta_\mathrm{c}$, which in general depends on the underlying gravity theory.  Up to now, the majority of analytic and numerical techniques calculate $\delta_\mathrm{c}$ within the framework of general relativity.  In this work, using simple physical arguments we estimate semi-analytically the PBH formation threshold within the framework of quantum gravity,  working for concreteness within loop quantum cosmology (LQC).  In particular,  for low mass PBHs formed close to the quantum bounce, we find a reduction in the value of $\delta_\mathrm{c}$ up to $50\%$ compared to the general relativistic regime quantifying for the first time to the best of our knowledge how quantum effects can influence PBH formation within a quantum gravity framework.  Finally, by varying the Barbero-Immirzi parameter $\gamma$ of loop quantum gravity (LQG) we show its effect on the value of $\delta_\mathrm{c}$ while using the observational/phenomenological signatures associated to ultra-light PBHs,  namely the ones affected by LQG effects, we propose the PBH portal as a novel probe to constrain the potential quantum nature of gravity. 

}


\begin{document}

\maketitle
\section{Introduction}
\label{sec:intro}
PBHs, firstly proposed in the early '70s ~\cite{1967SvA....10..602Z, Carr:1974nx,1975ApJ...201....1C}, form in the early universe, typically during the Hot Big Bang (HBB) radiation-dominated (RD) era out of the gravitational collapse of enhanced cosmological perturbations. According to recent arguments, PBHs can naturally account for a part or the totality of dark matter~\cite{Carr:2020xqk,Green:2020jor}, seed the large-scale structures through Poisson fluctuations~\cite{Meszaros:1975ef,Afshordi:2003zb,1984MNRAS.206..315C, Bean:2002kx} as well as the primordial magnetic fields through the presence of disks around them~\cite{Safarzadeh:2017mdy,Papanikolaou:2023nkx}. At the same time, they are associated with a plethora of gravitational-wave (GW) signals from black-hole merging events~\cite{Nakamura:1997sm, Ioka:1998nz, 
Eroshenko:2016hmn,Zagorac:2019ekv, Raidal:2017mfl} up to primordial scalar induced GWs~\cite{Bugaev:2009zh, Saito_2009, Nakama_2015, Papanikolaou:2020qtd,Domenech:2020ssp,Papanikolaou:2022chm} (for a recent 
review see \cite{Domenech:2021ztg}).  In particular, through the aforementioned GW portal, PBHs can act as well as a novel probe shedding light on the underlying gravity theory~\cite{Papanikolaou:2021uhe,Papanikolaou:2022hkg}. Other hints in favor of PBHs can be found here~\cite{2018PDU....22..137C}.

In the standard PBH formation scenario,  where PBHs form from the collapse of local overdensity regions, the PBH formation threshold $\delta_\mathrm{c}$ depends in general on the shape of the energy density perturbation profile of the collapsing overdensity~\cite{Germani:2018jgr,Musco:2018rwt,Escriva:2019phb,Musco:2020jjb,Escriva:2020tak} as well as on the equation-of-state parameter at the time of PBH gravitational collapse~\cite{Carr:1975qj,Harada:2013epa,Escriva:2020tak,Papanikolaou:2022cvo}.  This critical threshold value is very important since it can affect significantly the abundance of PBHs, a quantity which is constrained by numerous observational probes~\cite{Carr:2020gox}. 

From a historic perspective,  after a first analytic calculation of  $\delta_\mathrm{c}$  by B. Carr and S. Hawking in 1975 ~\cite{1974MNRAS.168..399C,Carr:1975qj}, $\delta_\mathrm{c}$ was studied mostly through numerical hydrodynamic simulations by~\cite{1978SvA....22..129N,1979ApJ...232..670B,1980SvA....24..147N,Niemeyer:1997mt,Shibata_1999}. Within the last decade,  there has been witnessed a remarkable progress regarding the determination $\delta_\mathrm{c}$ both at the analytic as well as at the numerical level.  In particular,  at the analytic level, T.Harada, C-M. Yoo $\&$ K. Kohri (HYK) in 2013~\cite{Harada:2013epa} refined the PBH formation threshold value obtained by Carr in 1975 by comparing the time at which the pressure sound wave crosses the overdensity collapsing to a PBH with the onset time of the gravitational collapse.  Their expression for $\delta_\mathrm{c}$ in the uniform Hubble gauge reads as:
\beq\label{delta_c-HYK}
\delta_\mathrm{c}=\sin^2\left(\frac{\pi\sqrt{w}}{1+3w}\right).
\eeq

At this point, it is very important to stress  that very recently there was exhibited a rekindled interest in the scientific community regarding the effect of non-linearities~\cite{Kawasaki:2019mbl,Young:2019yug,DeLuca:2019qsy,Germani:2019zez,Young:2020xmk} and non-Gaussianities~\cite{Young:2013oia,Young:2015cyn, Franciolini:2018vbk,Yoo:2019pma,Kehagias:2019eil} on the value of $\delta_\mathrm{c}$. In addition,  some first research works were also performed regarding the dependence of the PBH formation threshold on non sphericities~\cite{Kuhnel:2016exn,Yoo:2020lmg}, on anisotropies~\cite{Musco:2021sva}, on the velocity dispersion of the collapsing matter~\cite{Harada:2022xjp} as well as within the context of modified theories of gravity~\cite{Chen:2019ueb}.

In this work, we study semi-analytically the effect of the potential quantumness of spacetime on the determination of the PBH formation threshold by using simple physical arguments studying whether the PBH portal can act as a novel way to probe the quantum nature of gravity.  For concreteness, we work within the framework of loop quantum cosmology (LQC), being a symmetry-reduced model of loop quantum gravity (LQG)~\cite{Rovelli:1997yv,Ashtekar:2013hs}, which actually constitutes a nonperturbative and background-independent quantization of general relativity.  Very interestingly,  LQC is able to solve the problem of past and future singularities~\cite{Singh:2009mz} and provide the initial conditions for inflation, solving in this way naturally the flatness and the horizon cosmological problems~\cite{Ashtekar:2009mm}. It can also account for the large scale structure formation~\cite{Bojowald:2006zb} as well as for the currently observed cosmic acceleration~\cite{Wu:2008db,Chen:2008ca,Fu:2008gh}.  PBHs were studied firstly within the context of LQC in~\cite{Dwivedee:2011fd} where the PBH evolution was explored accounting for the effects of Hawking radiation and accretion in a LQC background.  In the present work, we investigate the effect of LQC on the PBH formation process and in particular at the level of the determination of the PBH formation threshold.

The paper is organized as follows: In \Sec{sec:LQG} we revise the basics of LQG and LQC. Then, in \Sec{sec:PBH_threshold_LQG} we determine semi-analytically the PBH formation threshold $\delta_\mathrm{c}$ by comparing the gravity and the sound wave pressure forces.  Followingly, in \Sec{sec:results} we present our results while \Sec{sec:conclusions} is devoted to conclusions. 

\section{The fundamentals of loop quantum gravity/cosmology}\label{sec:LQG}
Loop quantum gravity brings conceptually together the two fundamental pillars of modern physics, namely General Relativity (GR) and Quantum Mechanics (QM).  It constitutes actually a non-perturbative and background-independent quantization of general relativity~\cite{Ashtekar:2004eh,Thiemann:2007pyv}. In particular,  it is based on a connection-dynamical formulation of GR defined on a spacetime manifold $M=\mathbb{R}\times \Sigma$, where $\Sigma$ stands for the 3D spatial manifold.  
\subsection{The classical dynamics}\label{sec:LQG_classical}
Working within the Hamiltonian framework, the classical phase space consists of the Ashtekar-Barbero variables which are actually the two canonically conjugate variables of the theory.  These variables are the densitized triad $E^a_i$ and the Ashtekar connection $A^i_a$ defined as follows~\cite{Ashtekar:2004eh,Ashtekar:2003hd,Han:2005km,Thiemann:2007pyv}:
\begin{eqnarray}
E^a_i & = |\mathrm{det}(e^b_j)|^{-1}e^a_i,\\
A^i_a & = \Gamma^i_a + \gamma K^i_a,
\end{eqnarray}
where $e^a_i$ is the triad field, $\Gamma^i_a$ is the spin connection, $K^i_a$ is the extrinsic curvature and $\gamma$ is the so-called Barbero-Immirzi parameter which allows the quantisation procedure to be performed on a compact group.  Such a setup is based on a 3+1 decomposition of the metric written in the following form:
\beq
\mathrm{d}s^2 = N^2\mathrm{d}t^2 - q_{ab}(\mathrm{d}x^a + N^a\mathrm{d}t)(\mathrm{d}x^b + N^b\mathrm{d}t),
\eeq
where $q_{ab}=e^i_ae_{ib}$ is the spatial metric, $N$ is the lapse function and $N^i$ is the shift vector. 
This metric choice is in fact necessary in order to perform a Hamiltonian analysis of the theory. The latter relies on defining conjugate momenta of some variables (here the  3D–metric $q_{ab}$ and hence, requires defining a “formal” time variable)\footnote{Here, we conventionally denote as our temporal coordinates the ones perpendicular to the 3D spatial slices. This notation is convenient but does not preassume a preferred time.  As a consequence, general covariance is conserved.}. 
However, as said before, the LQG background equations will not depend on the choice of the spacetime  metric. This independence of the background on the choice of the spacetime foliation is associated to some constraints which can be derived performing a Dirac constraint analysis of the gravitational system. Firstly, one has the diffeomorphism constraint which renders the theory independent of the choice of the spatial geometry, i.e. of the shift vector, and secondly the Hamiltonian constraint which ensures the theory to be invariant under the choice of temporal coordinates, i.e. of the lapse function. These two constraints conserve the general spacetime covariance of the theory.  Thirdly, one gets the Gaussian constraint which makes the theory invariant under any rotations of the triad fields. Imposing therefore the aforementioned three constraints, one can derive the dynamical behavior of the theory.

At the classical level, the two canonically conjugate variables  $E^a_i$ and $A^i_a$ will be related with the following non-vanishing Poisson bracket:
\beq
\{A^i_a(\boldmathsymbol{x}),E^a_i(\boldmathsymbol{y})\} = 8\pi G\gamma \delta^b_a\delta^i_j\delta^{(3)}(\boldmathsymbol{x}-\boldmathsymbol{y}),
\eeq
while the dynamics of the theory will be governed by the following Hamiltonian acting on the canonical variables~\cite{Thiemann:1996aw,Yang:2015zda}:
\beq\label{eq:H_LQG}
H[N] =\frac{1}{8\pi G}\int_\Sigma \mathrm{d}^3\boldmathsymbol{x} N \left[ F^j_{ab} - (1+\gamma^2)\epsilon_{jmn}K^m_aK^n_b\right]\frac{\epsilon_{jkl}E^a_kE^b_l}{\sqrt{q}},
\eeq
where $F^j_{ab}$ is the curvature of the Ashtekar connection defined as $F^j_{ab}\equiv \partial_a A^j_b - \partial_bA^j_a + \epsilon^{ijk}A^j_aA^k_b$.

Working now within the spatially flat Friedman-Lema\^itre-Robertson-Walker (FLRW) model,  one introduces a fiducial cell $\mathcal{V}$ connected to a fiducial metric $^{o}q_{ab}$ and a fiducial orthonormal triad and co-triad $(^{o}e^a_i, ^{o}\omega^i_a)$ such as $^{o}q_{ab} = ^{o}\omega^i_{a}$ $^{o}\omega^i_b$. At the end, the reduced Ashtekar connection and densitized triad read as~\cite{Ashtekar:2003hd}
\beq
A^i_a = c V_0^{-1/3}{^{o}\omega^i_a}, \quad E^b_i = pV_0^{-2/3}\sqrt{\mathrm{det}(^oq)}{^{o}e^b_i},
\eeq
where $V_0$ is the fiducial volume as measured by the fiducial metric $^{o}q_{ab}$ and $c,p$ are functions of the cosmic time $t$.

In order to identify an internal clock of our theory, we introduce a dynamical massless scalar field described by the Hamiltonian:
\beq
H_\phi = \frac{p^2_\phi}{2|p|^{3/2}}.
\eeq
At the end, the cosmological classical phase space is composed of two congugate pairs $(c,p)$ and $(\phi,p_\phi)$ which obey the following Poisson brackets:
\beq
\{c,p\}=\frac{8\pi G}{3}\gamma, \quad \{\phi,p_\phi\}=1.
\eeq
with $|p|=a^2V^{2/3}_0$ and $c = \gamma \dot{a}V^{1/3}_0$.
Finally,  using he Hamiltonian constraint one obtains the usual Friedmann equation within GR for a flat FRLW model,
\beq
H^2 = \frac{8\pi G}{3}\rho.
\eeq
\subsection{The quantum dynamics}\label{sec:LQG_quantum}
Working now at the quantum level,  the classical phase space variables and the classical Hamiltonian will be promoted to quantum operators while the Poisson brackets will be replaced by commutation relations.  However, within quantum field theory, the commutation algebra of quantum operators requires integration over the 3D space, thus assuming a well pre-defined background. Nevertheless, this setup cannot be applied within the framework of LQG since we want a background independent theory.  For this reason, the quantisation process is performed at the level of two new canonical variables, namely the holonomy of the Ashtekar connection $h_e(A)$ along a curve $e\subset \Sigma$ and the flux of the densitized triad $F_S(E)$ along a 2-surface $S$ defined as~\cite{Ashtekar:2003hd}
\beq
h_e(A) \equiv \mathcal{P}exp\left(\int_e \tau_iA^i_a\mathrm{d}x^a\right), \quad F_S(E)\equiv \int_S\tau_iE^i_an^a\mathrm{d}^2y,
\eeq
where $\tau = -i\sigma_i/2$ ($\sigma_i$ are the Pauli matrices) with $[\tau_i,\tau_j] =\epsilon_{ijk}\tau^k$, $n^a$ is the unit vector vertical to the surface $S$ and $\mathcal{P}$ is a path-ordering operator. These functions constitute non-trivial SU(2) variables satisfying a unique holonomy-flux Poisson algebra~\cite{Ashtekar:2012cm,Engle:2016zac,Lewandowski:2005jk,Fleischhack:2004jc}.

Working within this representation one can  then construct a kinematical Hilbert space for the gravity sector which is actually the space of the square integrable functions on the Bohr compactification of the real line, i.e.   $\mathcal{H}^{\mathrm{grav}}_{\mathrm{kin}}\equiv L^2(R_\mathrm{Bohr},\mathrm{d}\mu_\mathrm{Bohr})$~\cite{Ashtekar:2003hd}. Regarding the matter sector, the respective kinematical Hilbert space is defined like in the standard Schrondigner picture as $\mathcal{H}^{\mathrm{matter}}_{\mathrm{kin}}\equiv L^2(R,\mathrm{d}\mu)$.  Thus, the whole kinematical Hilbert space of the theory is defined as  $\mathcal{H}_{\mathrm{kin}}\equiv \mathcal{H}^{\mathrm{grav}}_{\mathrm{kin}}\otimes \mathcal{H}^\mathrm{matter}_{\mathrm{kin}}$.

Focusing now on the homogeneous and isotropic FLRW model, usually denoted Loop Quantum Cosmology (LQC) and following the conventional quantisation $\bar{\mu}$ scheme~\cite{Ashtekar:2006wn} one introduces two new conjugate variables defined as follows:
\beq
u \equiv 2\sqrt{3}\mathrm{sgn}(p)/\bar{\mu}^3,\quad b\equiv \bar{\mu}c,
\eeq
where $\bar{\mu}=\sqrt{\Delta/|p|}$ and $\Delta = 4\sqrt{3}\pi\gamma G\hbar$ being the minimum nonzero eigenvalue of the area operator~\cite{Ashtekar:2008zu}. Finally, one can show that the new variables obey the following Poisson bracket:
\beq
\{b,u\}=\frac{2}{\hbar},
\eeq
and that in $\mathcal{H}^{\mathrm{grav}}$ there are two elementary operators, namely $\widehat{e^{ib/2}}$ and $\hat{u}$ related to the holonomy and  the flux conjugate variables.  In particular, it turns out that the eigenstates $\ket{u}$ of $\hat{u}$ form an orthonormal basis in $\mathcal{H}^{\mathrm{grav}}_{\mathrm{kin}}$ and the actions of these two operators in this basis can read as 
\beq
\widehat{e^{ib/2}}\ket{u}= \ket{u+1}, \quad \hat{u}\ket{u} = u\ket{u}.
\eeq
Letting now $\ket{\phi}$ being the orthonormal bases in $\mathcal{H}^{\mathrm{matter}}_{\mathrm{kin}}$ one can define $\ket{u,\phi} \equiv \ket{u}\otimes\ket{\phi}$ as the generalized basis of the whole kinematic Hilbert space $\mathcal{H}_{\mathrm{kin}}$.  Thus, after defining the relevant Hilbert space and the associated orthonormal basis, one can promote the Hamiltonian to a quantum operator.  In particular, it is possible to define a quasi-classical sharped initial state living in $\mathcal{H}_{\mathrm{kin}}$, which can be viewed as wavepacket around a classical trajectory. Consequently, expressing the Hamiltonian (\ref{eq:H_LQG}) in terms of fluxes and holonomies one can derive the expectation value of the Hamiltonian operator over the initial semi-classical sharped state which at the end will contain first order quantum corrections. Finally, accounting only for the holonomy correction\footnote{As it was shown from detailed numerical~\cite{Ashtekar:2006wn} and analytic~\cite{Corichi:2011rt,Corichi:2011sd} studies of the semi-classical states in the flat $k=0$ case, the 1st order quantum-corrected background equation \eqref{eq:Friedmann_LQG} gives us a good description of the quantum dynamics only when the volume of the universe is large at the bounce with respect to the Planck units. On the other hand, if the universe bounce occurs near the Planck scale, where inverse volume corrections become important, its dynamics is not well captured by the effective theory. Thus, we can safely neglect the inverse-volume corrections working in the regime where the effective quantum-corrected description of the background dynamics can be trusted. }
\beq\label{eq:Friedmann_LQG}
H^2 = \frac{8\pi G}{3}\rho\left(1-\frac{\rho}{\rho_\mathrm{c}}\right),
\eeq
where $\rho_\mathrm{c} = \frac{4\sqrt{3}\Mp^4}{\gamma^3}$.
As it can be seen from \Eq{eq:Friedmann_LQG} for $\rho>\rho_\mathrm{c}$ there is no physical evolution since $H^2<0$. One then finds that the effect of holonomies leads to a non-singular evolution where the classical Big Bang singularity is replaced by a non-singular quantum bounce where $\rho=\rho_\mathrm{c}$ and $H=0$. This bouncing point constitutes a transitioning point between a contracting ($H<0$) and an expanding phase ($H>0$).

At this point, it is important to stress that the aforementioned LQC background dynamics can be reproduced as well starting from a covariant quantum corrected effective field theory (EFT) action within modified gravity setups where the metric and the connection are regarded as independent~\cite{Olmo:2008nf}, being the LQG or another fundamental gravity theory.  Within metric gravity theories there have been some interesting attempts~\cite{Date:2008gq,Sotiriou:2008ya}, which however have not been successful in describing the regime where the non-perturbative quantum gravitational effects become significant, while at the same time the respective covariant effective actions are not in general uniquely defined potentially involving higher order curvature invariant terms~\cite{Ashtekar:2011ni}.  In addition, it is not evident on how one can treat anisotropic cosmological models within the EFT approach~\cite{Sotiriou:2008ya}.  

These EFTs can actually be treated as the low energy limit of the underlying fundamental theory, being the LQG or another fundamental bouncing gravity theory.  In this sense, the Barbero-Immirzi parameter $\gamma$ can be viewed as a fundamental parameter of another bouncing fundamental theory other than LQG.  However,  given the aforementioned limitations of the EFT approach, we will treat in the following the Barbero-Immirzi parameter $\gamma$ as the free fundamental parameter of the underlying gravity theory in the context of LQG.

\section{The threshold of primordial black hole formation  in loop quantum gravity}\label{sec:PBH_threshold_LQG}
Having introduced before the fundamentals of LQG, we estimate in this section the PBH formation threshold $\delta_\mathrm{c}$ accounting for effects of loop quantum gravity at the level of the background cosmic evolution.

To do so, we assume that the collapsing overdensity region is described by a homogeneous core (closed Universe) described by the following  fiducial metric:
\beq\label{overdense region metric}
\mathrm{d}s^2 = - \mathrm{d}t^2 + a^2(t)\left(\mathrm{d}\chi^2+\sin^2\chi \mathrm{d}\Omega^2\right),
\eeq
where $\mathrm{d}\Omega^2$ is the line element of a unit two-sphere and $a(t)$ is the scale factor of the perturbed overdensity region.  

For this type of closed homogeneous and isotropic spacetime foliations one can show that following the procedure as described in \Sec{sec:LQG} the modified Friedmann equation in $k=1$ LQC accounting only for the holonomy corrections \footnote{As in the case of the $k=0$ case, it was observed numerically that in the curvature-based $k=1$ theory as well~\cite{Ashtekar:2006es}, the effective quantum-corrected description of the background can be trusted only when the fiducial cell at the bounce is large in Planck units. Thus, the inverse-volume corrections, which are important in the ``small" fiducial cell limit, can be safely neglected in the regime where one trusts the effective quantum-corrected background dynamics description.} will read as
~\cite{Ashtekar:2006es,Corichi:2011pg}:
\beq\label{Friedmann equation - overdense region}
H^2 = \left(\frac{\dot{a}}{a}\right)^2 = \frac{1}{3\Mp^2}\left(\rho-\rho_*\right)\left[1- \frac{\rho - \rho_*}{\rho_\mathrm{c}}\right],
\eeq
where $\rho$ is the energy density of the overdense region and $\rho_*= \rho_\mathrm{c}\left[(1+\gamma^2)D^2 + \sin^2D\right]$ with $D\equiv\lambda (2\pi^ 2)^{1/3}/v^{1/3}$,  $\lambda^2 = 4\sqrt{3}\pi\gamma\ell^2_\mathrm{Pl}$~\cite{Motaharfar:2021gwi}, $\ell_\mathrm{Pl}$ being the Planck length and $v = 2\pi^2a^3$ being the physical volume of the unit sphere spatial manifold~\cite{Motaharfar:2021gwi}. Since $v$ increases with time one can expand \Eq{Friedmann equation - overdense region} in the limit $v\gg 1$~\cite{Ashtekar:2006es}. At the end,  keeping terms up to $\mathcal{O}(1/v^{2/3})$ one can show that \Eq{Friedmann equation - overdense region} takes the following form:
\beq\label{Friedmann equation-k=1}
H^2 = \left(\frac{\dot{a}}{a}\right)^2 = \frac{1}{3\Mp^2}\left[\rho\left(1-\frac{\rho}{\rho_\mathrm{c}}\right) - \frac{3\Mp^2}{a^2}\left(1 - 2\frac{\rho}{\rho_\mathrm{c}}\right)\right] = \frac{F(\rho)}{3\Mp^2} -\frac{G(\rho)}{a^2},
\eeq
where $F(\rho) = \rho(1-\rho/\rho_\mathrm{c})$ and $G(\rho) = 1- 2\rho/\rho_\mathrm{c}$.  In the limit where $\gamma = 0$, $\rho_\mathrm{c}\rightarrow\infty$ and one recovers the standard GR $k=1$ Friedmann equation $H^2  = \frac{\rho}{3\Mp^2} - \frac{1}{a^2}$.

Working now with the background, the latter will behave as the standard homogeneous and isotropic FLRW background whose fiducial metric reads as
\beq\label{background metric}
\mathrm{d}s^2 = - \mathrm{d}t^2 + a^2_\mathrm{b}(t)\left(\mathrm{d}r^2+r^2\mathrm{d}\Omega^2\right)
\eeq
and whose modified Friedmann equation within LQC will read as
\beq\label{Friedmann equation - background universe}
H^2_\mathrm{b} = \left(\frac{\dot{a}_\mathrm{b}}{a_\mathrm{b}}\right)^2 = \frac{\rho_\mathrm{b}}{3\Mp^2}\left(1-\frac{\rho_\mathrm{b}}{\rho_\mathrm{c}}\right).
\eeq

In this setup, the collapsing overdense region corresponds to the region where $0\leq \chi\leq \chi_a$ and the areal radius at the edge of the overdensity will read as
\beq
R_\mathrm{a} = a\sin\chi_\mathrm{a}.
\eeq

At this point, we need to stress that the characteristic size of the overdensity is initially super-horizon and will reenter the cosmological horizon when the areal radius of the overdensity becomes equal to the cosmological horizon $H^{-1}$, i.e.
\beq\label{eq:hc_1}
\frac{1}{H_\mathrm{hc}} = a_\mathrm{hc}\sin\chi_a,
\eeq
where the index ``hc" denotes quantities at the horizon crossing time.  Writing now the energy density of the overdensity as $\rho = \rho_\mathrm{b}(1+\delta)$, where $\delta\equiv  \frac{\delta\rho}{\rho_\mathrm{b}}$, one can plug $\rho$ into \Eq{Friedmann equation-k=1} and working within the uniform Hubble gauge where $H = H_\mathrm{b}$ they can recast \Eq{eq:hc_1} as
\beq\label{eq:hc_time_chi_a}
\sin^2\chi_a = \frac{1}{G^\delta_\mathrm{hc}}\left(\frac{F^\delta_\mathrm{hc}}{F_\mathrm{hc}}-1\right),
\eeq
where $F^\delta_\mathrm{hc} = F\left[\rho_\mathrm{b,hc}(1+\delta)\right]$ and $G^\delta_\mathrm{hc} = G\left[\rho_\mathrm{b,hc}(1+\delta)\right]$. 

Once then the overdensity region enters in causal contact with the rest of the Universe,  i.e. when its chatacteristic scale crosses the cosmological horizon, it will initially follow the cosmic expansion and at some point it will detach from it starting to collapse to form a black hole horizon. This basically happens at the time of maximum expansion of the overdensity, when the Hubble parameter in \Eq{Friedmann equation - background universe} becomes zero, i.e. $H_m=0$, or equivalently when 
\beq\label{eq:a_m}
a_\mathrm{m}=\frac{3\Mp^2 G_m}{F_\mathrm{m}},
\eeq
with the subscript ``m" denoting quantities at the maximum expansion time.

Having derived above the horizon crossing time and the time at maximum expansion we establish below a criterion for PBH formation by investigating the necessary conditions for the triggering of the gravitational collapse process.  Doing so, we confront the gravitational force which pushes matter inwards and enhances in this way the black hole gravitational collapse with the sound wave pressure force which pushes matter outwards, thus disfavoring the collapse of the overdensity. In particular, the criterion which we adopt is the requirement that the time at which the pressure sound wave crosses the radius of the overdensity region should be larger than the time at the maximum expansion, which is actually the time of the onset of the gravitational collapse.  Thus, the sound pressure force will not have time to disperse the collapsing fluid matter to the background medium and prevent in this way the collapsing process. Equivalently,  we require that the proper size of the overdensity $\chi_a$ is larger than the sound crossing distance by the time of maximum expansion $\chi_\mathrm{s}$, i.e.
\beq
\chi_a > \chi_\mathrm{s}.
\eeq

To compute now the sound crossing distance by the time of maximum expansion we assume matter in terms of a perfect fluid characterized by a constant equation-of-state (EoS) parameter $w$, defined as the ratio between the pressure $p$ and the energy density $\rho$ of the fluid, $w\equiv p/\rho$. 

It is important to stress here that one may become confused from the fact that the effective LQC Friedmann equations in the $k=0$ and the $k=1$ used to describe here the background and the overdensity respectively rely on the existence of a massless scalar field~\cite{Taveras:2008ke}. Thus,  one would legitimately ask how can we describe matter within the three-zone model in terms of a perfect fluid characterized by an EoS parameter $w$. In order to answer this question,  let us highlight that the  effective LQC Friedmann equations describing the background dynamics were also derived assuming matter in form of dust~\cite{Willis:2004br} as well as in form of a massive scalar field~\cite{Singh:2006im,Calcagni:2008ig} while in~\cite{Mielczarek:2012uzm} they were derived by not assuming a specific matter Hamiltonian.  In \Sec{sec:LQG_quantum}, we just used a massless scalar field as an internal clock to derive the LQC corrected background dynamics.  The effective background LQC Friedmann equations \eqref{eq:Friedmann_LQG} and \eqref{Friedmann equation-k=1} are actually independent on the matter content of the Universe. The situation is different when one goes to the perturbation level.  For more details see~\cite{Mielczarek:2012uzm}.

Thus, having clarified the effect of the Universe's matter content on the form of the background LQC Friedmann equations, one can write the sound wave propagation equation for a Universe filled with a perfect fluid characterised by an EoS parameter $w$ as
\beq
a\frac{\mathrm{d}\chi}{\mathrm{d}t}=\sqrt{w},
\eeq
where we used the fact that for a perfect fluid with a constant EoS parameter the square sound wave $c^2_\mathrm{s}$ is equal to $w$, i.e.  $c^2_\mathrm{s}=w$. At the end,  $\chi_\mathrm{s}$ can be recast in the following form:
\beq\label{eq:chi_s}
\chi_\mathrm{s}  = \sqrt{w}\int_{t_\mathrm{ini}}^{t_\mathrm{m}}\frac{\mathrm{d}t}{a} = \frac{\sqrt{w}}{3}\int_{\rho_\mathrm{b,m}}^{\rho_\mathrm{ini}}\frac{\mathrm{d}\rho_\mathrm{b}}{(1+w)\rho_\mathrm{b}\sqrt{\left(\frac{\rho_\mathrm{b,m}}{
\rho_\mathrm{b}}\right)^{\frac{2}{3(1+w)}}\frac{G_\mathrm{m}F(\rho_\mathrm{b})}{F_\mathrm{m}}-G(\rho_\mathrm{b})}},
\eeq
where in the last equality we transformed $t$ to $\rho_\mathrm	{b}$ accounting for the fact that for a Universe filled with a perfect fluid $\rho_\mathrm{b}\propto a^{-3(1+w)}$ and $a\propto t^{\frac{2}{3(1+w)}}$. In addition, we used as well \Eq{eq:a_m} in order to express $a_\mathrm{m}$ in terms of $F_\mathrm{m}$ and $G_\mathrm{m}$. 

At the end, using \Eq{eq:hc_time_chi_a} the criterion for PBH formation reads as
\beq\label{eq:PBH_criterion}
\frac{1}{G^\delta_\mathrm{hc}}\left(\frac{F^\delta_\mathrm{hc}}{F_\mathrm{hc}}-1\right)>\sin^2\chi_\mathrm{s}.
\eeq

To determine therefore the PBH formation threshold, one can follow the following procedure: From \Eq{eq:hc_time_chi_a}, one should firstly determine the ratio $\rho_\mathrm{hc}/\rho_\mathrm{m}$ for a given value of $\chi_a$ and then solve numerically the inequality \Eq{eq:PBH_criterion} in order to extract the value of the critical energy density contrast $\delta_\mathrm{c}$ required for the overdensity region to collapse and form a PBH.  

Practically,  one should compute $\chi_\mathrm{s}$ from \Eq{eq:chi_s} for a given value of $\rho_\mathrm{m}$ and equate $\chi_\mathrm{s}$ with $\chi_a$. Then, solving \Eq{eq:hc_time_chi_a} they will extract the ratio $\rho_\mathrm{hc}/\rho_\mathrm{m}$ and then plugging it into \Eq{eq:PBH_criterion} they can extract numerically $\delta_\mathrm{c}$.  At the end, given the fact that $\rho_\mathrm{b,hc}<\rho_\mathrm{c}$, i.e.  PBHs form after the quantum bounce, and that $\delta<1$ since we want to be within the perturbative regime,  one can show from \Eq{eq:hc_time_chi_a}  that
\beq\label{eq:delta_c_estimate_LQG}
\delta_\mathrm{c} \simeq \sin^2\chi_\mathrm{s},
\eeq
with $\chi_\mathrm{s}$ given by \Eq{eq:chi_s}.

At this point, we should stress that the above expression for the value of $\delta_\mathrm{c}$ is a lower bound estimate of its true value since it assumes the homogeneity of the collapsing overdensity region which in general is not the case when one is met with strong pressure gradients.  Thus,  it is strictly valid for regimes where $w\ll 1$. In the case of GR, it was shown that \Eq{eq:delta_c_estimate_LQG} gives a reliable estimate for the value of the threshold  even in the case where $w=1/3$ when compared with the results from numerical simulations~\cite{Harada:2013epa}. However, PBH formation was never studied before in a rigorous way within the context of LQG through numerical simulations. To that end, \Eq{eq:delta_c_estimate_LQG} can be seen as a first reliable estimate for the value of $\delta_\mathrm{c}$ giving us a qualitative tendency for the production of the small mass PBHs formed close to the quantum bounce to be significantly enhanced within LQC. To get however an accurate answer on the true value of $\delta_\mathrm{c}$ one needs to run high cost hydrodynamic simulations within LQC, something which was never performed before in the literature to the best of our knowledge and which goes beyond the scope of the present work.


\section{Results}\label{sec:results}
\begin{figure}[h!]
\begin{center}
\includegraphics[width=0.495\textwidth]{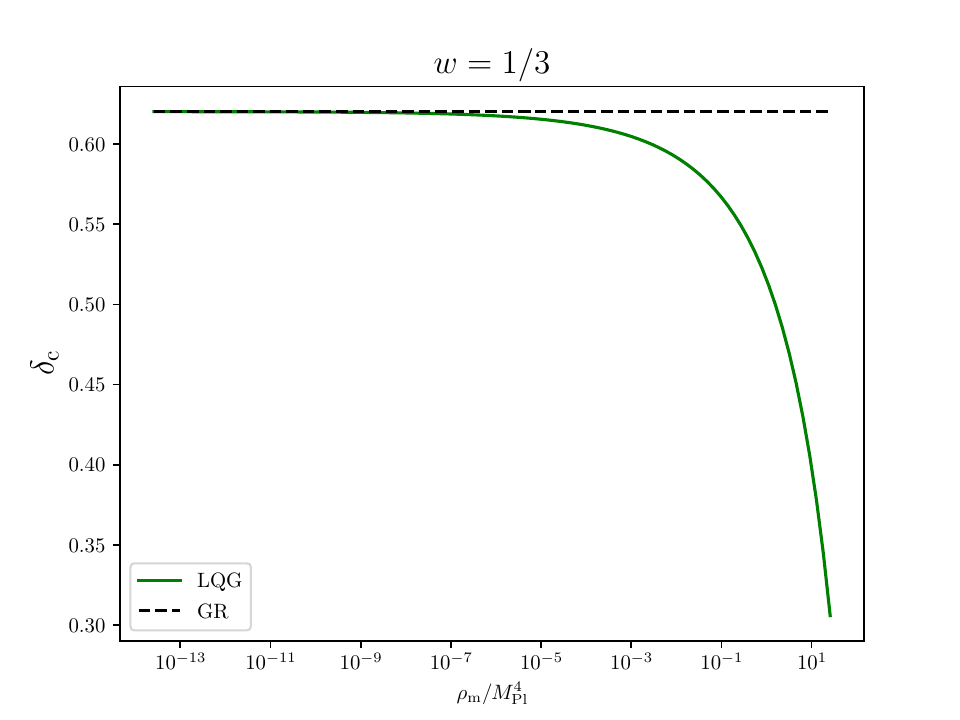}
\includegraphics[width=0.495\textwidth]{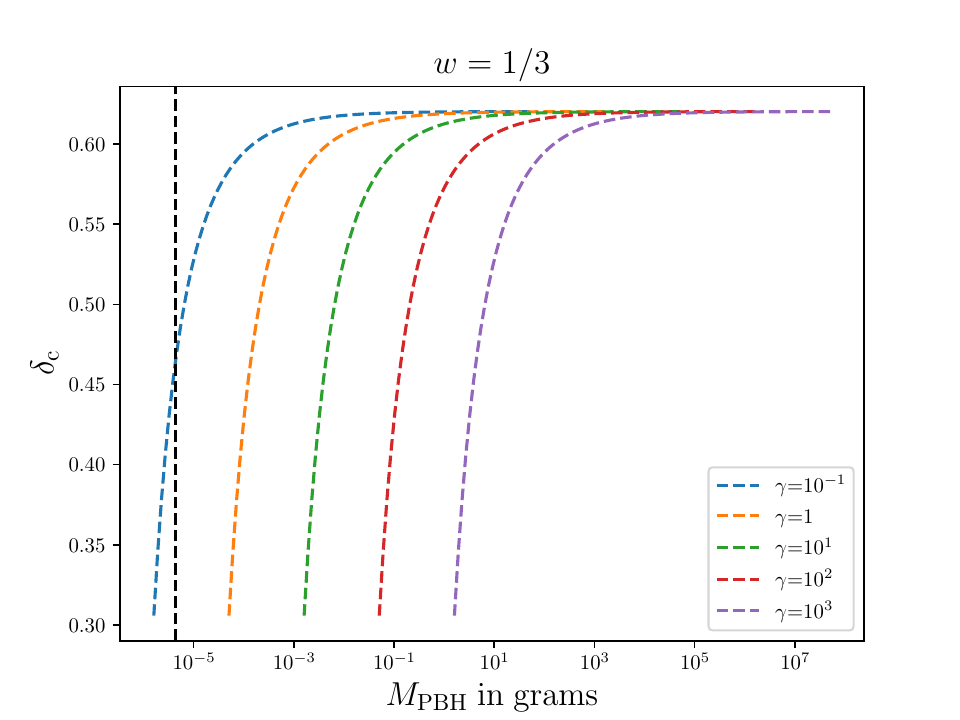}
\caption{{\it{Left Panel:The PBH formation threshold in the uniform Hubble gauge in the radiation-dominated era ($w=1/3$) as a function of the energy density at the onset of the  PBH gravitational collapse in LQG (green curve) and in GR (\Eq{delta_c-HYK}) (black dashed curve). Right Panel: The PBH formation threshold in the uniform Hubble gauge in the radiation-dominated era ($w=1/3$)  as a function of the primordial black hole mass $M_\mathrm{PBH}$ in LQG for different values of the Barbero-Immirzi parameter $\gamma$.  The vertical black dashed line corresponds to $M_\mathrm{PBH}=\Mp$.  }}}
\label{fig:delta_c_LQG_vs_GR}
\end{center}
\end{figure}

Following the procedure described above, we calculate here the PBH formation threshold $\delta_\mathrm{c}$ within the framework of LQG and we compare it with its value in GR.  In particular, in the left panel of \Fig{fig:delta_c_LQG_vs_GR} we show the PBH formation threshold as a function of the energy density at the time of maximum expansion $\rho_\mathrm{m}$ by fixing the EoS parameter $w=1/3$ since we study PBH formation during the RD era and the value of the Barbero-Immirzi parameter $\gamma=0.2375$ obtained from the computation of the entropy of black holes~\cite{Meissner:2004ju}.  Interestingly, we see a deviation from GR for high energies at the time of maximum expansion which correspond to very small mass PBHs forming close to the quantum bounce. This behavior is somehow expected since in this high energy regime, one expects to see a quantifiable effect of the quantum nature of gravity. In particular,  we observe a drastic reduction of the value of $\delta_\mathrm{c}$ in this region of high values of $\rho_\mathrm{m}$ up to $50\%$ compared to the GR case.  \footnote{The horizontal axis of the left panel is expressed in terms of the reduced Planck mass $\Mp$ which is related to the Planck mass $m_\mathrm{Pl}$ as $\Mp^2=m^2_\mathrm{Pl}/(8\pi)$.  Thus,  one can go to values of $\rho_\mathrm{m}$ up to $\rho_\mathrm{c}/10= (\frac{4\sqrt{3}\Mp^4}{\gamma^3})/10 \simeq 51.7\Mp^4 \simeq 0.082m^4_\mathrm{Pl}<m^4_\mathrm{Pl}$ since $\rho_\mathrm{hc}\sim 10\rho_\mathrm{m}$ and $\rho_\mathrm{hc}\leq \rho_\mathrm{c}$ as stated above \Eq{eq:N_max_expansion}. }

Let us stress here that we considered in \Fig{fig:delta_c_LQG_vs_GR} PBH formation during a RD era. This RD era is expected to appear as a short pre-inflationary stage in LQC inflationary setups due to gravitational particle production occuring during the bounce phase~\cite{Graef:2020qwe,SVicente:2022ebm} or within LQC bouncing cosmological setups without inflation where an RD era occurs soon after the onset of the expanding phase~\cite{Corichi:2007am,Bojowald:2008pu,Amoros:2013nxa}. In this way, inflation if there exists does not have time to ``wash out" the quantum corrections on the background dynamics and one is able to see an enhanced PBH production in LQC. 

This reduction of $\delta_\mathrm{c}$ should be related with a smaller cosmological/sound horizon in LQC compared to GR as it can be speculated from \Eq{eq:delta_c_estimate_LQG}. To see this,  let us find the necessary conditions to get a cosmological horizon in LQC smaller than that in GR.  Doing so, one should require that 
\beq\label{eq:cosmological_horizon_criterion}
H^2_\mathrm{LQC}>H^2_\mathrm{GR}\Leftrightarrow \frac{\rho}{3\Mp^2}\left(\rho-\frac{\rho}{\rho_\mathrm{c}}\right) -\frac{1}{a^2}\left(1-2\frac{\rho}{\rho_\mathrm{c}}\right)>\frac{\rho}{3\Mp^2} -\frac{1}{a^2}\Leftrightarrow \rho <\frac{6\Mp^2}{a^2}.
\eeq
For $\rho=\rho_\mathrm{m}$ and $a=a_\mathrm{m}$ as given by \Eq{eq:a_m} one can verify that the inequality \ref{eq:cosmological_horizon_criterion} is identically satisfied. Thus, indeed the cosmological horizon in LQC is smaller than in GR leading to a reduction of $\delta_\mathrm{c}$ compared to its GR value.

At this point, it is important to highlight that the PBH formation threshold depends as well, as already mentioned in the Introduction,  on the shape of the of power spectrum of the cosmological perturbations which collapsed to form PBHs~\cite{Germani:2018jgr}, which can either lower or raise $\delta_\mathrm{c}$. In particular,  a very narrow-peaked power spectrum tends to raise the threshold and potentially can counterbalance the $50\%$ reduction of $\delta_\mathrm{c}$ within LQC.  This should be carefully investigated by extracting the effect of the shape of the power spectrum of cosmological perturbations on $\delta_\mathrm{c}$ within LQC, something which has not yet been studied in the literature and goes beyond the scope of this work. However, in general one expects a reduction in $\delta_\mathrm{c}$ within LQC compared to the case of GR, unless one works with a particularly narrow matter power spectrum.

Then,  we consider the Barbero-Immirzi parameter  $\gamma$ as a free parameter of the underlying quantum theory in the context of LQG. In particular,  despite the fact that the Bekenstein-Hawking entropy has been standardly used as a way to fix the value of $\gamma$, the dependence of the entropy calculation on $\gamma$ is controversial, and the value $\gamma \simeq 0.2375$,  calculated using thermodynamical arguments, is not broadly accepted~\cite{Engle:2010kt,Bianchi:2012ui,Wong:2017pgl}. In fact,  the choice to vary this parameter is motivated by the fact that $\gamma$ is actually a coupling constant with a topological term in the gravitational action, with no consequence at the level of the classical equations of motion~\cite{Asante:2020qpa,Perlov:2020cgx,Broda:2010dr,Mercuri:2009zt,Pigozzo:2020zft,Carneiro:2020uww}. Thus, we vary the Barbero-Immirzi parameter within the range of $0.1<\gamma<1000$ accounting for observational constraints for the duration of inflation after a quantum bounce~\cite{Benetti:2019kgw,Barboza:2022hng}.  At the end, we plot in the right panel of \Fig{fig:delta_c_LQG_vs_GR}  the PBH formation threshold $\delta_\mathrm{c}$ as a function of the PBH mass for different values of the parameter $\gamma$ within the observationally allowed range $\gamma\in[0.1,1000]$.  We set the lower bound on the PBH mass equal to the Planck mass as predicted within the quantum gravity approach~\cite{Coleman:1991ku} (See vertical black dashed line in the right panel of \Fig{fig:delta_c_LQG_vs_GR}).

In order to get the PBH mass, we account for the fact that the PBH mass is of the order of the cosmological horizon mass at horizon crossing time.  Solving at the end numerically \Eq{eq:hc_time_chi_a} we found that  $\rho_\mathrm{hc}/\rho_\mathrm{m}\sim 10$. This corresponds to 
\beq\label{eq:N_max_expansion}
N = \ln\left(\frac{a_\mathrm{m}}{a_\mathrm{hc}}\right) = \frac{1}{4}\ln\left(\frac{\rho_\mathrm{hc}}{\rho_\mathrm{m}}\right)\sim 0.6\mathrm{\;\mathrm{e-folds}}
\eeq
passing from horizon crossing time up to the onset of the gravitational collapse process at $\rho=\rho_\mathrm{m}$,  thus being in agreement with the results from PBH numerical simulations~\cite{Musco:2004ak}.  As expected, when we increase the value of $\gamma$ the overall mass range moves to higher masses given the fact that higher values of $\gamma$ are equivalent with lower values of $\rho_\mathrm{c}$, thus the quantum bounce happens at later times. Consequently, PBHs if formed will form at later times, thus will acquire larger masses. 

Interestingly, independently on the value of the Barbero-Immirzi variable $\delta_\mathrm{c }$ is reduced on the low mass region, which for $\gamma<1000$ corresponds to masses $M_\mathrm{PBH}<10^3\mathrm{g}$\footnote{Recently, it was shown in~\cite{Choudhury:2023vuj,Choudhury:2023jlt} that one-loop corrections to the renormalized primordial power spectrum rules out the possibility of having large mass PBHs within single-field inflationary models making the low-mass PBHs with $M_\mathrm{PBH}<10^3\mathrm{g}$ very well motivated from the theoretical point of view.  These very light PBHs can be naturally produced in LQG setups as discussed in~\cite{Modesto:2009ve}.}. In particular, this  reduction in $\delta_\mathrm{c}$ in this very small PBH mass range will be equivalent with an enhancement in their abundances, entailing in this way tremendous consequences on the phenomenology associated to ultra-light PBHs. Indicatively, we mention here that these ultra-light PBHs can trigger early PBH-matter dominated eras~\cite{Inomata:2019zqy,Papanikolaou:2020qtd,Domenech:2020ssp} before BBN and produce the DM relic abundance and the hot Standard Model (SM) plasma~\cite{Lennon:2017tqq} reheating in this way the Universe through their evaporation~\cite{Martin:2019nuw}. Furthermore, they can account for the Hubble tension through the injection to the primordial plasma of light dark radiation degrees of freedom~\cite{Hooper:2019gtx,Nesseris:2019fwr} while at the same time they can produce naturally the baryon assymetry through CP violating out-of-equilibrium decays of their Hawking evaporation products~\cite{Barrow:1990he,Bhaumik:2022pil,Gehrman:2022imk}.   Consequently,  one can constrain the above mentioned observational/phenomenological signatures by studying PBH formation within the context of LQG while vice-versa given the above mentioned phenomenology one can constrain the Barbero-Immirzi parameter $\gamma$ which is the fundamental parameter within LQG.  In this way,  PBHs are promoted as a novel probe to constrain the potential quantum nature of gravity.

\section{Conclusions}\label{sec:conclusions}

PBHs firstly introduced in the '70s are of great significance, since they can naturally account for a part or all of the dark matter sector, while at the same time they might seed the formation of large-scale structures through Poisson fluctuations. Moreover, they can also offer the seeds for the progenitors of the black-hole merging events recently detected by LIGO/VIRGO as well as for the supermassive black holes present in the galactic centers. Their formation was mainly studied within the context of general relativity using both analytic and numerical techniques.

In this work, we studied PBH formation within the context of LQC by investigating the impact of the potential quantum character of spacetime on the critical PBH formation threshold $\delta_\mathrm{c}$,  whose value can crucially affect  the abundance of PBHs, a quantity which is constrained by numerous observational probes.  In particular, by comparing the gravitational force with the sound wave pressure force during the process of the gravitational collapse we obtained a reliable estimate on the value of $\delta_\mathrm{c}$.

Interestingly, we found that for low mass PBHs formed close to the quantum bounce,  the value of $\delta_\mathrm{c}$ is drastically reduced up to $50\%$ compared to the general relativistic regime with tremendous consequences for the observational/phenomenological footprints of such small PBH masses. In this way, we quantified for the first time to the best of our knowledge how quantum effects can influence PBH formation in the early Universe within a quantum gravity framework.

Finally,  by treating the Barbero-Immirzi parameter $\gamma$ as the free parameter of LQG we varied its value by studying its effect on the value of the PBH formation threshold.  As expected, we found an overall shift of the PBH masses affected by the choice of $\gamma$. Very interestingly, we showed as well that using the observational and phenomenological signatures associated to ultra-light PBHs, namely the ones affected by LQG effects,  one can  constrain the quantum parameter $\gamma$.  At this point, we should highlight the fact that our formalism can be applied to any quantum theory of gravity giving an explicit form for the equations of the background cosmic evolution establishing in this way the PBH portal as a novel probe to constrain the potential quantum nature of gravity.

\section*{Acknowledgments}
The author acknowledges financial support from the Foundation for Education and European Culture in Greece as well the contribution of the COST Action
CA18108 ``Quantum Gravity Phenomenology in the multi-messenger approach''.
\bibliographystyle{JHEP} 
\bibliography{PBH}
\end{document}